\documentclass[11pt]{article}

\usepackage[paperwidth=8.5in,paperheight=11in,portrait,top=1in,bottom=1.in,left=1.15in,right=1.15in]{geometry}
\usepackage[utf8]{inputenc}
\usepackage{authblk}

\usepackage{amsfonts,amsmath,amssymb,amsthm}
\usepackage{latexsym,mathrsfs,mathtools,bm}
\usepackage{graphicx,subcaption,epsfig,caption,float,xcolor}
\usepackage{enumitem}
\usepackage{chngcntr}

\usepackage{tikz}
\usetikzlibrary{calc}

\usepackage[hidelinks]{hyperref}
\usepackage{bookmark}

\theoremstyle{plain}
\newtheorem{thm}{Theorem}[section]

\newtheorem{prop}[thm]{Proposition}

\numberwithin{equation}{section}

\newcommand{\om}{\overline{m}}
\newcommand{\op}{\overline{p}}

\newcommand{\cA}{\mathcal{A}}
\newcommand{\cB}{\mathcal{B}}
\newcommand{\cC}{\mathcal{C}}
\newcommand{\cV}{\mathcal{V}}

\newcommand{\II}{\mathbb{I}}
\newcommand{\ZZ}{\mathbb{Z}}

\numberwithin{equation}{section}

\title{\bf Bethe ansatz diagonalization of the Heun--Racah operator}

\renewcommand*{\Affilfont}{\normalsize\small}
\author[1]{Pierre-Antoine Bernard}
\author[2]{Gauvain Carcone}
\author[3]{Nicolas Cramp\'e}
\author[4]{Luc Vinet\vspace{.5em}}

\affil[1,2,4]{Centre de Recherches Math\'ematiques, Universit\'e de Montr\'eal, \newline\vspace{.9em}
P.O. Box 6128, Centre-ville Station, Montr\'eal (Qu\'ebec), H3C 3J7, Canada.}
\affil[3]{Institut Denis-Poisson CNRS/UMR 7013 - Universit\'e de Tours - Universit\'e d'Orl\'eans, \newline\vspace{.9em}
Parc de Grandmont, 37200 Tours, France.}
\affil[4]{IVADO, Montr\'eal (Qu\'ebec), H2S 3H1, Canada. \newline\vspace{.9em}}

{
	\makeatletter
	\renewcommand\AB@affilsepx{: \protect\Affilfont}
	\makeatother
	\affil[ ]{E-mail addresses}
	\makeatletter
	\renewcommand\AB@affilsepx{, \protect\Affilfont}
	\makeatother
	\affil[1]{pierre-antoine.bernard@umontreal.ca}
	\affil[2]{gauvain.carcone@umontreal.ca}
	\affil[3]{crampe1977@gmail.com}
	\affil[4]{vinet@crm.umontreal.ca}
}

\begin{document}

\date{\today} 
\maketitle

\begin{abstract}

The Heun--Racah operator is diagonalized with the help of the modified algebraic Bethe ansatz. This operator is the most general bilinear expression in two generators of the Racah algebra.
A presentation of this algebra is given in terms of dynamical operators, and allows the construction of Bethe vectors for the Heun--Racah operator.
The associated Bethe equations are derived for both the homogeneous and inhomogeneous cases.
\end{abstract}

\maketitle

\vspace{3mm}

\section{Introduction}

The differential Heun operator defines the Fuchsian second order differential equation with four regular singularities. It generalizes the Jacobi operator
which gives rise to the hypergeometric equation with three regular singularities. This standard differential Heun operator is recovered from the general bilinear combination of the Jacobi differential
operator and the operator multiplication by the variable \cite{GVZ}. This construction can be extended to any pair of operators defining
a bispectral problem and leads to the definition of algebraic Heun operators \cite{GVZ2}. The polynomials of the Askey scheme are bispectral: they obey a differential or difference equation and satisfy a 3-term recurrence relation. The associated algebraic Heun operator is designated by the name of the family. The Heun--Jacobi operator is thus identified with the standard one \cite{VZ,BTVZ,CVZ,BCTVZ}.
The Heun operators have found many applications in various contexts. For example, they were seen in \cite{GVZ2,BCV3,BCV4} to provide an algebraic interpretation
for the existence of the commuting operators in the time and band limiting problem \cite{Sle,Lan}.
They also naturally lead to the commuting tridiagonal operator \cite{Eis1} arising in the computation of the entanglement entropy for 1D free Fermion models
\cite{CNV,CNV2} with couplings given by the recurrence coefficients of orthogonal polynomials of the Askey scheme.
This has been generalized to free fermions on vertices of distance regular graphs \cite{CGV,BCV,BCV2} or to multidimensional networks
associated to the multivariate Krawtchouk polynomials \cite{BCNPPV}.

In this paper, we focus on the Heun--Racah operator. The Racah polynomials are at the top of the Askey scheme and provide through different limits all the discrete hypergeometric polynomials.
Its recurrence and difference relations satisfy the Racah algebra introduced in the algebraic study of the $6j$-symbol and of the Racah problem for $\mathfrak{su}(2)$ \cite{GrZh,GeVZ}
(see \cite{CGPV} for a review). The Racah algebra can be embedded in $sl_2$ \cite{GZ,Koo,BH,CSV}, is associated to the DAHA $(C_1^\vee,C_1)$ \cite{Hua}
and appears in the context of superintegrable models \cite{KKM,Pos,KMP,GeVZ2,GeVZ3}. Its finite dimensional irreducible representations have been classified in \cite{HB2}.
The Heun--Racah operator and its associated algebra have been studied specifically in \cite{BCTVZ}.

It is known that the spectrum of an algebraic Heun operator has no analytical expression. The algebraic Bethe ansatz, a method introduced to solve quantum integrable system \cite{STF}, has been used to study the Heun--Askey--Wilson operator and to express its eigenvalues in terms of the Bethe roots \cite{BasP}. This approach has been applied to the Heun operators of Lie type \cite{BCSV}.
These computations are based on the fact that the Heun operator is identified in a transfer matrix of a quantum integrable system, namely the XXZ spin chain with generic boundaries
for the Heun--Askey--Wilson type, and the Gaudin model for the one of Lie type. The Racah algebra has not been found however in the context of quantum integrable systems.
Let us notice that the Hahn algebra, that lies between the Racah algebra and the Lie ones, appears in this integrable context and has been related to the XXX spin chain and the Yangian \cite{CRVZ}.
Since the Racah algebra seems not to be directly related to an integrable model, a generalization of these methods is required to study the Heun--Racah operator with the Bethe ansatz.
In \cite{BCSV}, a direct method to construct the necessary objects for the Bethe ansatz has been applied successfully to study the spectrum of a Heun operator associated to
a time and band limiting problem. We show here that this direct method can be also be adopted to diagonalize the Heun--Racah operator by a Bethe ansatz approach.

The plan of this paper is as follows. In section \ref{secracah}, a presentation of the Racah algebra $R(3)$ in terms of dynamical operators is introduced. Representations of $R(3)$ and the Heun--Racah operator are given in section \ref{heunrep}. Section \ref{heundiag} covers the diagonalization by the modified algebraic Bethe ansatz. In particular, Bethe vectors are defined in terms of the dynamical operators and the associated Bethe equations are given for both the homogeneous inhomogeneous cases.

\section{Racah algebra and dynamical operators} \label{secracah}

The Racah algebra $R(3)$ is generated by $X$, $Y$ and $Z$ satisfying the following defining relations
\begin{eqnarray}
&&[X,Y]=Z,\label{eq:r1}\\
&&[Z,X]=-2 X^2-2 \{X,Y\}+bX+d_2,\label{eq:2}\\
&&[Y,Z]=-2 Y^2-2 \{X,Y\}+bY+d_1,\label{eq:r3}
\end{eqnarray}
where $\{X,Y\}=XY+YX$ and $b$, $d_1$ and $d_2$ are defining parameters. 

For an arbitrary parameter $\rho \in \mathbb{C}$, let us introduce a set of operators labeled by two additional real parameters $u$ and $m$ 
\begin{eqnarray}
 \cA(u,m)&=&\frac{1}{2m\rho-1}\big(g_0(u,m)+g_1(u,m)X+g_1(-1,m)Y+Z+\rho \{X,Y\}\big), \label{eq:AXY}\\
 \cB(u,m)&=&f_0(u,m)+f_1(u,m)X+f_1(-1,m)Y+2m Z+\{X,Y\},\label{eq:BXY}\\
 \cC(u,m)&=& \cB(u,-m+1/\rho),\label{eq:CXY}
\end{eqnarray}
where
\begin{eqnarray}
f_0(u,m)&=&\frac{(4m^2-1)(u^2 -  2b -1)}{8}-d_2,\quad f_1(u,m)=\frac{4m^2-u^2}{2},\\
 g_0(u,m)&=&\rho f_0(u,m)+ (2m\rho-1)\left( \frac{(4m-u+1)(2b+1-u^2)}{8} -\frac{d_1-d_2}{u-1} \right) ,\\
 g_1(u,m)&=&\frac{(u-2m)(2m\rho-\rho u-2)}{2}.
\end{eqnarray}

The choice of operators \eqref{eq:AXY}-\eqref{eq:CXY} has been dictated by the fact that 
the defining relations of the Racah algebra \eqref{eq:r1}-\eqref{eq:r3} give the following relations between $\cA(u,m)$, $\cB(u,m)$ and $\cC(u,m)$:
\begin{eqnarray}
 \cB(u,m+1)\cB(v,m)&=&\cB(v,m+1)\cB(u,m),\label{eq:BB}\\
 \cA(u,m)\cB(v,m)&=&k_1(u,v)\cB(v,m)\cA(u,m-1)\nonumber\\
 && +\cB(u,m)\big( k_2(u,v,m) \cA(v,m-1)+k_2(u,-v,m)\cA(-v,m-1) \big), \label{eq:AB} \\
 \cC(v,m)\cA(u,m)&=&k_1(u,v)\cA(u,m-1)\cC(v,m)\nonumber\\
 && +\big( k_2(u,v,m) \cA(v,m-1)+k_2(u,-v,m)\cA(-v,m-1) \big)\cC(u,m),
\end{eqnarray}
with functions $k_1$ and $k_2$ given by
\begin{eqnarray}
&&  k_1(u,v)=\frac{(u-2)^2-v^2}{u^2-v^2},\qquad
 k_2(u,v,m)=\frac{ (v-1)( \rho(u- v-4 m)+2)}{v(v-u)(2m\rho-1)}.
\end{eqnarray}

Operators verifying similar relations are usually called \textit{dynamical operators} in the context of the algebraic Bethe ansatz.
Such operators have been introduced in \cite{Bax, FT} to study the XYZ model; the approach was then adapted to analyze the $XXZ$ spin chain 
 with boundary fields that are not parallel \cite{Cao,BellP}. Up to then, these dynamical operators have been constructed from the $R$-matrix formalism.
 However, as explained in the introduction, the Racah algebra, in contrast to the Askey-Wilson algebra, does not appear in this context.
 The idea of this paper consists in constructing the dynamical operators without using an $R$-matrix.

The dynamical operators  $\cA(u,m)$, $\cB(u,m)$ and $\cC(u,m)$ provide a new presentation of the Racah algebra $R(3)$ as stated in the following proposition
\begin{prop}
 Both relations \eqref{eq:BB} and \eqref{eq:AB} with the expressions \eqref{eq:AXY} and \eqref{eq:BXY} are equivalent to the defining relations \eqref{eq:r1}-\eqref{eq:r3} of the Racah algebra $R(3)$.
\end{prop}

\proof The $R(3)$-relations imply the relations  \eqref{eq:BB} and \eqref{eq:AB} since the dynamical operators have been constructed for this purpose. 
To prove the other implication, one replaces the dynamical operators by their expressions \eqref{eq:AXY} and \eqref{eq:BXY} in relations \eqref{eq:BB} and shows that the coefficients of $u,v$ and $m$ imply the $R(3)$-relations.
\endproof

\section{Heun operator and representation} \label{heunrep}

The Heun-Racah operator is the most general bilinear operator in terms of $X$ and $Y$. It reads explicitly as follows
\begin{equation}
 W=r_0+r_1X +r_2 Y +r_3 XY + r_4YX, \label{eq:W}
\end{equation}
for $r_i$ some free parameters.
It has been shown in \cite{NT} that it is the most general operator tridiagonal in the two bases where either $X$ or $Y$ is diagonal.

Representations of the Racah algebra $R(3)$ are provided by the recurrence and difference relations of the Racah polynomials $R_n(x(x+\gamma + \delta + 1); \alpha, \beta, \gamma, \delta)$ \cite{Koek,Zhe1}. 
These are obtained from the action of the following $(N+1)\times (N+1)$-matrices on $(N+1)$-dimensional vectors with entry $x$ given by one of polynomial evaluated at $x$ \cite{Koek}:
\begin{equation} 
X=\frac{(\alpha+\beta)(\alpha+\beta+2)}{4}\II_{N+1}+\begin{pmatrix}
 -B(0)-D(0)   & D(1)\\
  B(0)  & -B(1)-D(1)  & D(2) \\
    && \ddots & \\
    && B(N-1)& -B(N)-D(N) 
   \end{pmatrix}\,, \label{eq:Xm}
\end{equation}
and 
\begin{equation} 
Y=\begin{pmatrix}
    \lambda_0 &\\
    & \lambda_1 & \\
    && \ddots \\
    &&& \lambda_N
   \end{pmatrix}\,,\label{eq:Ym}
\end{equation}
with $\alpha=-N-1$, $N\in \ZZ_{>0}$ and 
\begin{eqnarray}
&&B(x)=\frac{(x+\alpha+1)(x+\beta+\delta+1)(x+\gamma+1)(x+\gamma+\delta+1)}{(2x+\gamma+\delta+1)(2x+\gamma+\delta+2)},\\
&&D(x)=\frac{x(x-\alpha+\gamma+\delta)(x-\beta+\gamma)(x+\delta)}{(2x+\gamma+\delta)(2x+\gamma+\delta+1)},\\
&& \lambda_x=\frac{(2x+\gamma+\delta+2)(2x+\gamma+\delta)}{4}.
\end{eqnarray}
By abuse of language, the same symbol will denote the abstract generators and the above matrices. 
From now on, $X$ and $Y$ stand for the matrices \eqref{eq:Xm} and \eqref{eq:Ym}. 
They satisfy relations \eqref{eq:r1}-\eqref{eq:r3} with 
\begin{eqnarray}
 && b =  (\beta+\delta)(\beta-\gamma)+(\alpha-\delta)(\alpha-\gamma)+\delta^2+\gamma^2-2  , \\
 && d_1 =\frac{(\gamma^2-\delta^2)(2\beta-\gamma+\delta)(\gamma+\delta-2\alpha)}{8},\qquad d_2 = \frac{(\alpha^2-\beta^2)(\alpha-\beta-2\delta)(2\gamma-\beta-\alpha)}{8}.
\end{eqnarray}
This realization gives an expression for the Heun operator \eqref{eq:W} as a tridiagonal matrix. 
In the following section, a modified algebraic Bethe ansatz, based on the dynamical operators $\cA(u,m)$, $\cB(v,m)$ and $\cC(u,m)$, is used to investigate its spectrum. 

\section{Diagonalization of the Heun operator}\label{heundiag}

\paragraph{Heun operator.} For latter convenience, the Heun operator is rewritten as follows 
\begin{eqnarray}
W=-2\frac{\rho}{\rho-1} XY  +s_1 X+\frac{s_2^2-1}{2\rho(\rho-1)} Y +[X,Y], \label{eq:WW}
\end{eqnarray}
where $\rho$, $s_1$ and $s_2$ are free parameters. We recover the generic Heun operator \eqref{eq:W}, which has five free parameters, by performing an affine transformation of the operator  \eqref{eq:WW}.
This Heun operator has a simple expression in terms of the dynamical operator $\cA$ given by
\begin{eqnarray}
 W=  h_1(u) \cA(u,\om)+h_1(-u) \cA(-u,\om)  +h_2(u), \label{eq:WA}
\end{eqnarray}
where 
\begin{eqnarray}
 &&\om=\frac{s_2-\rho +1}{2\rho},\\
 &&h_1(u)=\frac{s_1}{2u}-\frac{(\rho u-\rho+s_2)^2-1}{4\rho u(\rho-1)},\qquad h_2(u)=-\frac{h_1(u)g_0(u,\om)+h_1(-u)g_0(-u,\om)}{2 \om \rho-1}.
\end{eqnarray}

\paragraph{Bethe vectors.} 
The Bethe vectors are defined by
\begin{equation}
 | x_1,x_2,\dots,x_p;m\rangle =\cB(x_1,m)\cB(x_2,m-1)\dots \cB(x_p,m-p+1)|0\rangle, \label{eq:BV}
\end{equation}
for $p\geq 1$ and
\begin{equation}
 |0\rangle= (1,0,\dots,0)^t.
\end{equation}
It is worth noting that the Bethe vector does not depend on the ordering of the $x_i$ because of the relation \eqref{eq:BB} between two operators $\cB$.
In the context of spin chains, the operator $\cB(x,m)$ is interpreted as the creation operator of a quasi-particle of momentum $x$. 
We shall show that these vectors become eigenvectors of the Heun operator $W$ if the parameters $x_i$ satisfy some relations called Bethe equations. In this case,
the parameter $x_i$ are called Bethe roots.

\paragraph{Action of $\cA$ on Bethe vectors.}
Using the commutation relations \eqref{eq:AB} between the operators  $\cA(u,m)$ and $\cB(u,m)$, the action of $\cA(u,m)$ on the Bethe vector \eqref{eq:BV} can be computed.
Indeed, one gets
\begin{align}
 &\cA(u,m)| x_1,x_2,\dots,x_p;m\rangle= \prod_{j=1}^p k_1(u,x_j)\cB(x_1,m)\cB(x_2,m-1)\dots \cB(x_p,m-p+1) \cA(u,m-p)|0\rangle \label{eq:ABV}\\
 &+  \sum_{\epsilon=\pm} \sum_{r=1}^p  k_2(u,{\epsilon}x_r,m)\prod_{\genfrac{}{}{0pt}{}{\ell=1}{\ell\neq r}}^p k_1({\epsilon}x_r,x_{\ell}) \cB(x_1,m)\dots \cB(u,m-r-1) \dots \cB(x_p,m-p+1) \cA(\epsilon x_r,m-p)|0\rangle.\nonumber
\end{align}
 
 To arrive at the previous relation, the specific representations of $X$ and $Y$ are not needed.
 They are however required to derive the action of $\cA$ on the vector $|0\rangle$. Namely, 
\begin{equation}
 \cA(u,m)|0\rangle=\xi(u,m)|0\rangle  +\zeta(u,m) \cB(u,m)|0\rangle,
\end{equation}
where
\begin{eqnarray}
 &&\zeta(u,m)=\frac{\delta\rho+\gamma\rho+2m\rho+2\rho-\rho u-2}{(2m\rho-1)(\delta+\gamma-2m+2-u)},\\
 &&\xi(u,m)=\frac{((u+N)^2 -(\beta-\gamma+\delta)^2)( \beta^2 -(u-N-2-\gamma-\delta)^2)(\delta+\gamma-2m+u)}{8(u-1)(\delta+\gamma-2m+2-u)}.
\end{eqnarray}

\paragraph{Action of $W$ on Bethe vectors.}
We are now in a position to determine the action of the Heun operator $W$ on the Bethe vector $\cV_p=|x_1,x_2,\dots,x_p;\om\rangle$.
Firstly, let us remark that 
\begin{equation}
 h_1(u)k_2(u,v,\om)+h_1(-u)k_2(-u,v,\om)=\frac{f_1(v)}{\rho(\rho-1)(u^2-v^2)}
\end{equation}
where 
\begin{equation}
f_1(v)=\big( 2\rho(\rho-1)s_1-(\rho v-\rho+s_2+1)(\rho v-\rho+s_2-1) \big)(1-1/v).
\end{equation}
The important feature of the previous relation is that $f_1(v)$ does not depend on $u$.
Secondly, using the expression \eqref{eq:WA} of $W$ in terms of $\cA$ and the previous action \eqref{eq:ABV} of $\cA$ on the Bethe vectors, one gets
\begin{align}
 W\cV_p &= w_p \cV_p   + \sum_{r=1}^p \frac{1}{\rho(\rho-1)(u^2-x_r^2)} \ U_r\ | x_1,\dots,x_{r-1},u,x_{r+1},\dots,x_p;\om\rangle \nonumber \\
& + \psi(u,p) \, \cB(x_1,\om)\cB(x_2,\om-1)\dots \cB(x_p,\om-p+1) \cB(u,\om-p)|0\rangle,\label{eq:WV}
\end{align}
where 
\begin{align}
 w_p &=h_1(u) \xi(u,\om) \prod_{j=1}^p k_1(u,x_j) + h_1(-u) \xi(-u,\om) \prod_{j=1}^p k_1(-u,x_j) + h_2(u),\\
   U_r &= \sum_{\epsilon=\pm} f_1(\epsilon x_r) \xi({\epsilon}x_r,\om-p) \prod_{\genfrac{}{}{0pt}{}{\ell=1}{\ell\neq r}}^p k_1({\epsilon}x_r,x_{\ell}),\\
  \psi(u,p)&=\frac{\rho^2\Big( 2(1-\rho)s_1+(\gamma+\delta+2+2p)(\delta \rho+\gamma\rho+2\rho(p+1)-2)\Big)}{(1-\rho)((\delta\rho+\gamma\rho+\rho (3+2p)-s_2-1)^2-\rho^2 u^2)} \nonumber\\
&\times\prod_{r=1}^p \frac{(\delta\rho+\gamma\rho+\rho(1+2p)-s_2-1)^2-\rho^2 x_r^2}{(\delta\rho+\gamma\rho+\rho(3+2p)-s_2-1)^2-\rho^2 x_r^2}\label{eq:psip}
\end{align}
The expression \eqref{eq:psip} has been obtained by factoring terms in
\begin{align}
\psi(u,p)&= \sum_{\nu=\pm }h_1(\nu u)\Biggl(\zeta(\nu u,\om-p)\prod_{j=1}^p k_1(\nu u,x_j) + \sum_{\epsilon=\pm} \sum_{t=1}^p \zeta(\epsilon x_t,\om-p)k_2(\nu u,\epsilon x_t,\om)\prod_{\genfrac{}{}{0pt}{}{\ell=1}{\ell\neq t}}^p k_1(\epsilon x_t,x_\ell)\Biggr).
\end{align}

At this point, there two cases to consider. In the first case, there exists a $\op$ such that $\psi(u,{\op})=0$. It is referred to as the homogeneous case and can be solved by the \textit{usual} algebraic Bethe ansatz.

In the second (inhomogeneous) case, one must deal with the last term in \eqref{eq:WV}, i.e. a vector of the form $\cB(x_1,m)\cB(x_2,m-1)\dots \cB(x_p,m-p+1) \cB(u,m-p)|0\rangle$. For $p=N$, there exists a formula (see equation \eqref{eq:maba}) which allows to absord these unwanted terms in the sum on the rhs of \eqref{eq:WV}. Such a relation is the foundation of the modified Bethe ansatz introduced in \cite{BC} for the XXX spin chain with generic boundaries, and later used in different contexts \cite{ABGP,BasP,BellP,BCSV,BCV4,Cra1,Cra2}. We explain both cases in detail in the following paragraphs.

\paragraph{Diagonalization of $W$ for the homogeneous case.}  In the first line of the expression \eqref{eq:psip}, there is a factor which is independent of $u$ and $x_r$. This factor vanishes if 
\begin{equation}
p=\frac{1-\gamma\rho-\delta\rho-2\rho\pm \sqrt{2s_1\rho^2-2s_1\rho+1}}{2\rho}.
\end{equation}
If the rhs of the above equation is an integer $\overline{p}$, a Bethe vectors such that $\psi(u,\op)=0$ can be constructed.
Relation \eqref{eq:WV} then reduces to 
\begin{align}
 W\cV_{\op} &= w_{\op} \cV_{\op}   + \sum_{r=1}^{\op }\frac{U_r}{\rho(\rho-1)(u^2-x_r^2)} \ | x_1,\dots,x_{r-1},u,x_{r+1},\dots,x_{\op};\om\rangle. \label{eq:WVh}
\end{align}
Finally $\cV_{\op}$ becomes an eigenvector if $U_r=0$ for any $r=1,\dots {\op}$. These equations are referred to as the Bethe equations and read explicitly as
\begin{equation}
\frac{f_1( x_r) \xi(x_r,\om-{\op})}{f_1(-x_r) \xi(-x_r,\om-{\op})}=- \prod_{\genfrac{}{}{0pt}{}{\ell=1}{\ell\neq r}}^{\op} \frac{(x_r+2)^2-x_{\ell}^2}{(x_r-2)^2-x_{\ell}^2}\,.
\end{equation}
Note that this case, where the Bethe ansatz method simplifies for 
a particular choice of the parameter, appears already in the study of the XXZ spin chain with generic boundaries \cite{Cao}.

\paragraph{Diagonalization of $W$ for the inhomogeneous case.}

In the inhomogeneous case, we consider Bethe vectors with $p=N$ Bethe roots. We recall that $N+1$ is the size of the matrices.
Therefore, in the second line of \eqref{eq:WV}, there are Bethe vectors with $N+1$ variables. 
Fortunately, such Bethe vectors can be expressed as linear combinations of Bethe vectors depending only on $N$ variables as follows
\begin{eqnarray}\label{eq:maba}
\displaystyle  | x_1,x_2,\dots,x_N,u;\om\rangle &=&\tau_u \ | x_1,x_2,\dots,x_N;\om\rangle  \\&+&\displaystyle \sum_{j=1}^{N} \frac{(\gamma+\delta-2\om+2N+2)^2-u^2}{x_j^2-u^2} \tau_j  | x_1,x_2,\dots,x_{j-1},u,x_{j+1}\dots,x_{N};\om\rangle,\nonumber
\end{eqnarray}
where 
\begin{eqnarray}
\tau_j&=&\frac{(2\om-N)^2-\beta^2}{8} \prod_{ \genfrac{}{}{0pt}{}{k=1}{k\neq j}}^{N}  
 \frac{(\gamma+\delta-2\om+2N+2)^2-x_k^2}{x_j^2-x_k^2}\prod_{k=0}^N (x_j^2-(\beta-\gamma+\delta-N+2k)^2)
\nonumber\\ &&\times \prod_{k=1}^N \frac{1}{(2\om-2\delta-\beta-N-2k)(2\om-2\gamma+\beta-N-2k)},\\
\tau_u&=&\frac{(2\om-N)^2-\beta^2}{8} \prod_{k=1}^{N}  \frac{(\gamma+\delta-2\om+2N+2)^2-x_k^2}{u^2-x_k^2}
\prod_{k=0}^N (u^2-(\beta-\gamma+\delta-N+2k)^2)\nonumber\\
&&\times \prod_{k=1}^N \frac{1}{(2\om-2\delta-\beta-N-2k)(2\om-2\gamma+\beta-N-2k)}.
\end{eqnarray}
This expression is proven by direct computation for $N=1,2,3,4$. It is conjectured for bigger $N$. A proof of a similar relation can be found in \cite{ABGP, Cra1,Cra2}.
Using the above relation, the action of $W$ on Bethe vectors given by \eqref{eq:WV} becomes 
\begin{align}
 W\cV_N &= (w_N+w^{(i)}_N) \cV_N   + \sum_{r=1}^N\frac{1}{\rho(\rho-1)(u^2-x_r^2)} \left(U_r+U_r^{(i)}\right)| x_1,\dots,x_{r-1},u,x_{r+1},\dots,x_N;\om\rangle \nonumber \\
\end{align}
where 
\begin{align}
&w^{(i)}_N= \tau_u \psi(u,N)\\
& U_r^{(i)}=\tau_r\rho\Big( 2(1-\rho)s_1+(\gamma+\delta+2+2N)(\delta \rho+\gamma\rho+2\rho(N+1)-2)\Big)\nonumber\\
&\times\prod_{q=1}^N \frac{(\delta\rho+\gamma\rho+\rho(1+2N)-s_2-1)^2-\rho^2 x_q^2}{(\delta\rho+\gamma\rho+\rho(3+2N)-s_2-1)^2-\rho^2 x_q^2}\label{eq:psip2}
\end{align}
These additional terms in the eigenvalues and the Bethe equations are referred as the inhomogeneous terms  (and indexed by $(i)$)
in reference to the similar terms appearing in the context of the XXZ spin chain \cite{Cao1,Cao2,Nepo}.

As in the homogeneous case, $\cV_N$ becomes an eigenvector if $U_r+U_r^{(i)} = 0$ for any $r = 1 \dots N$. These are the inhomogeneous Bethe equations, which read as 

\begin{equation}
\frac{f_1( x_r) \xi(x_r,\om-{N})}{f_1(-x_r) \xi(-x_r,\om-{N})}+\frac{U_r^{(i)}}{f_1(-x_r)\xi(-
x_r,\om-N)\displaystyle\prod_{\genfrac{}{}{0pt}{}{\ell=1}{\ell\neq r}}^N k_1(-x_r,x_{\ell})}=- \prod_{\genfrac{}{}{0pt}{}{\ell=1}{\ell\neq r}}^{N} \frac{(x_r+2)^2-x_{\ell}^2}{(x_r-2)^2-x_{\ell}^2}\,.
\end{equation}

\section{Conclusion}

Dynamical operators have been defined with relations providing an equivalent presentation of the Racah algebra. The Heun--Racah operator was given in terms of these dynamical operators and diagonalized with the Bethe ansatz. The resulting Bethe equations are homogeneous or inhomogeneous depending on the parameters defining the Heun operator. 

The method proposed here does not rely on the usual $R$--matrix presentation. It should be interestingto use the same approach to find sets of dynamical operators for the Hahn and the Bannai--Ito algebra so as to diagonalize the corresponding Heun operators. Indeed, for these two algebras the generic Heun operator does not seem to arise in the usual $R$-matrix formalism. Let us remark that while the Hahn algebra can be defined within this $R$-matrix formalism (see \cite{CRVZ}), the generic Heun operator with five parameters does not appear however. 

The usual method consists in defining the algebra with some non-dynamical operators thanks to relations (FRT relation, reflection equation) based on the $R$--matrix (see e.g. \cite{BasP}). Dynamical operators are then constructed as linear combinations of these operators.  It could also be instructive to find directly a presentation of the relations of these dynamical operators within a $R$--matrix formalism in order to find more complicated Heun operators.
For example, Heun operators associated to multivariate polynomials has been introduced in \cite{BCNPPV} and it should prove interesting to study their spectra through the algebraic Bethe ansatz.

\paragraph{Acknowledgements.}
PAB holds an Alexander-Graham-Bell scholarship from the Natural Sciences and Engineering Research Council of Canada (NSERC).
GC thanks the Department of Physics of the Université de Montréal for partial support.
NC thanks the CRM for its hospitality and are supported by the international research project AAPT of the CNRS and the ANR Project AHA ANR-18-CE40-0001. 
The research of LV is supported by a Discovery Grant from the Natural Sciences and Engineering Research Council (NSERC) of Canada.

\end{document}